\documentclass[12pt]{article}
\usepackage{geometry}

\usepackage{svg}
\usepackage{amsmath, amssymb, graphicx,color,mathtools,amsthm,xcolor}
\usepackage{caption, subcaption}
\usepackage{algorithm, algorithmic}
\usepackage{cite}
\usepackage{subcaption}
\usepackage{array}
\usepackage{url}
\usepackage{tikz}
\usepackage{tabularx}
\usepackage{microtype}
\usepackage{hyperref,color}

\definecolor{webgreen}{rgb}{0,.35,0}
\definecolor{webbrown}{rgb}{.6,0,0}
\definecolor{RoyalBlue}{rgb}{0,0,0.9}
\definecolor{purp}{rgb}{0.6,0.05,0.8}
\definecolor{ora}{rgb}{0.7,0.35,0.02}

\hypersetup{
   colorlinks=true, linktocpage=true, 
   urlcolor=webbrown, linkcolor=RoyalBlue, citecolor=webgreen,
   pdfauthor={Hangyu Li, Gary P. T. Choi},
   pdfsubject={Planar morphometry via functional shape data analysis and quasi-conformal mappings}
}

\title{Planar morphometry via functional shape data analysis and quasi-conformal mappings}
\author{Hangyu Li$^{1}$, Gary P. T. Choi$^{1,\ast}$\\
\\
\footnotesize{$^{1}$Department of Mathematics, The Chinese University of Hong Kong}\\
\footnotesize{$^\ast$To whom correspondence should be addressed; E-mail: ptchoi@cuhk.edu.hk}
}
\date{ }

\begin{document}

\maketitle

\begin{abstract}
The study of shapes is one of the most fundamental problems in life sciences. Although numerous methods have been developed for the morphometry of planar biological shapes over the past several decades, most of them focus solely on either the outer silhouettes or the interior features of the shapes without capturing the coupling between them. Moreover, many existing shape mapping techniques are limited to establishing correspondence between planar structures without further allowing for the quantitative analysis or modelling of shape changes. In this work, we introduce FDA-QC, a novel planar morphometry method that combines functional shape data analysis (FDA) techniques and quasi-conformal (QC) mappings, taking both the boundary and interior of the planar shapes into consideration. Specifically, closed planar curves are represented by their square-root velocity functions and registered by elastic matching in the function space. The induced boundary correspondence is then extended to the entire planar domains by a quasi-conformal map, optionally with landmark constraints. Moreover, the proposed FDA-QC method can naturally lead to a unified framework for shape morphing and shape variation quantification. We apply the FDA-QC method to various leaf and insect wing datasets, and the experimental results show that the proposed combined approach captures morphological variation more effectively than purely boundary-based or interior-based descriptions. Altogether, our work paves a new way for understanding the growth and form of planar biological shapes.
\end{abstract}

\section{Introduction}
The study of shapes plays a significant role in biology. For instance, to compare two biological structures, a fundamental task is to quantify their geometric differences. Also, understanding the developmental process of a biological structure often requires tracking changes in the size and shape of its different regions over time. The use of mathematical transformations and mappings for comparing different biological shapes and quantifying their geometric variation has been studied for over a century~\cite{d1917growth,kendall1984shape,bookstein1986size,zelditch2012geometric,bookstein2013measurement,gu2023classical}. With the advancement of technology for acquiring different biological datasets, there is an increasing need for the development of mathematical and computational tools for both \emph{morphometrics} (quantitative analysis of form) and \emph{morphogenesis} (developmental process of form). 

For planar biological structures, conformal mappings have emerged as a natural tool because they preserve angles and local shapes. In particular, conformal maps have been used to analyze planar plant leaves, quantify growth patterns, and study anisotropic expansion in developmental experiments~\cite{alim2016leaf,mitchison2016conformal,dai2022minimizing}. Related ideas have also been applied to insect wings, where conformal parameterizations provide a convenient canonical domain for comparing wing outlines and vein patterns~\cite{alba2021global}. As a generalization of conformal maps, quasi-conformal maps allow controlled shear and anisotropic distortion, and hence can better accommodate complex shape differences and the alignment of feature landmarks while still possessing low local geometric distortions. For this reason, quasi-conformal techniques have been used in recent works for various planar morphometry and mapping-based morphogenesis problems~\cite{jones2013planar,choi2018planar,mosleh2025data}. Besides, large deformation diffeomorphic mapping methods~\cite{joshi2000landmark,beg2005computing} have also been widely used in the computation of landmark-based mappings between planar shapes.

In parallel, there is a complementary line of work that represents planar shapes primarily through their boundary curves. In particular, many prior works have studied the development of shape metrics on planar curves~\cite{younes1998computable,mio2007shape}. Conformal welding, for example, describes a Jordan curve by a homeomorphism of the unit circle obtained by gluing two conformal maps along the boundary~\cite{sharon20062d}. This boundary-centric viewpoint leads naturally to functional representations and infinite-dimensional Riemannian geometry on spaces of curves. In particular, the square-root velocity function (SRVF) framework of functional shape data analysis (FDA) provides a powerful elastic representation of curves, from which optimal registrations and geodesic distances between planar curves can be computed efficiently in the function space~\cite{srivastava2010shape,srivastava2016functional}. FDA-based methods have been successfully applied to the comparison between closed biological outlines, the registration of developmental trajectories, and the construction of low-dimensional shape descriptors. More recently, Zhang and Srivastava~\cite{zhang2021elastic} developed an extension of the method to solid planar objects using tensor field representations. 

Despite these advances, existing approaches typically emphasize either the boundary or the interior, but rarely both in a unified framework. In particular, quasi-conformal mapping methods usually require a prescribed boundary correspondence to drive the mapping~\cite{choi2018planar,mosleh2025data}. For instance, the method in~\cite{choi2018planar} relies on prescribing boundary feature points on the source and target shapes as boundary landmarks, followed by a curvature-guided registration of every segment between them. Also, the method in~\cite{mosleh2025data} models the continuous growth of biological shapes using quasi-conformal flows but requires the boundary point correspondence at various time points. Identifying such correspondences manually is tedious and may be ambiguous for large-scale biological datasets with substantial shape variability. On the other hand, FDA-based curve registration methods provide mathematically optimal boundary correspondences but are intrinsically one-dimensional and do not directly describe how interior material points should be transported between domains. As a result, there is still a gap between elastic boundary registration and quasi-conformal interior mapping in planar morphometry.

\begin{figure}[t]
    \centering
    \includegraphics[width=\linewidth]{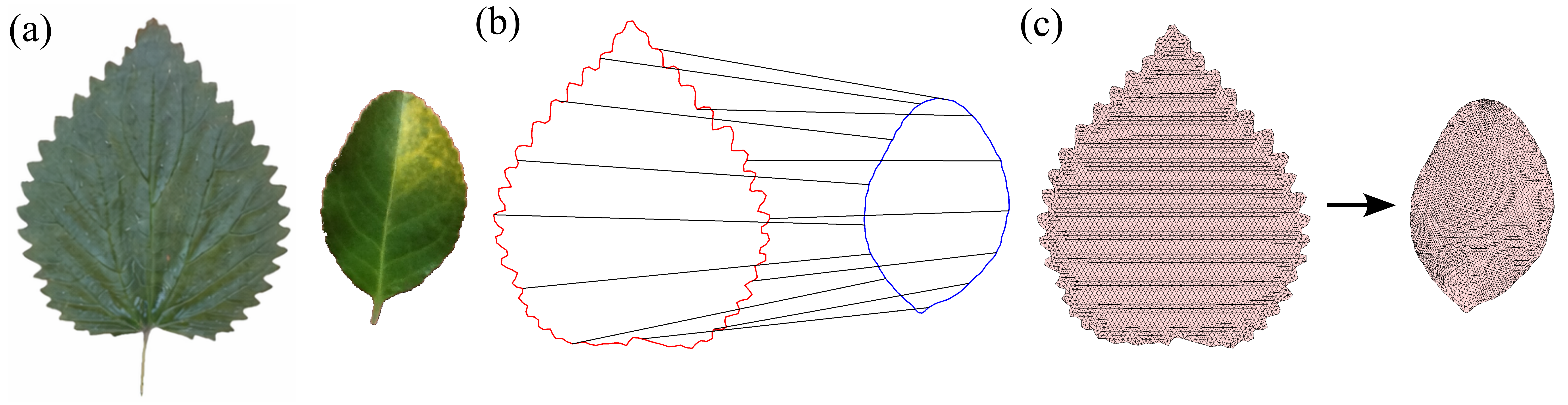}
\caption{\textbf{The proposed FDA-QC method integrating functional shape data analysis (FDA) and quasi-conformal (QC) mapping techniques for planar morphometry.} (a)~A \emph{Urtica dioica} leaf (left) and a \emph{Euonymus japonicus} leaf (right) adapted from~\cite{leafuci}. (b)~Boundary alignment achieved using the FDA technique, where homologous points along the two closed curves are matched through an optimal reparameterization. (c)~Planar correspondence between the two leaf shapes induced by the registered boundaries and interior QC mapping.}
    \label{fig:illustration_fdaqc}
\end{figure}

In this work, we develop FDA-QC, a novel mathematical method for morphometrics and morphogenesis of planar shapes that integrates functional data analysis and quasi-conformal geometry (Fig.~\ref{fig:illustration_fdaqc}). Specifically, we first
propose a method for computing planar mappings that can optimize both the boundary and interior shape correspondences, optionally with the presence of interior feature landmark constraints. Then, we extend the mapping method and further develop a continuous morphing process between two shapes. Finally, we utilize the mapping method to construct shape dissimilarity measures for quantifying the geometric variation in planar structures. As demonstrated by our experiments on different leaf and insect wing datasets, the proposed framework can be effectively applied to both morphometrics and morphogenesis of planar biological shapes.

The rest of this paper is organized as follows. In Section~\ref{sec:math_background}, we introduce the basic concepts of functional shape data analysis and quasi-conformal geometry. In Section~\ref{sect:main}, we describe our proposed FDA-QC mapping method and explain how the method can be further utilized for shape morphing and shape variation quantification. In Section~\ref{sect:experiment}, we present the experimental results with various biological datasets, demonstrating the effectiveness of our proposed framework. We conclude our work and discuss possible future directions in Section~\ref{sect:conclusion}.

\section{Mathematical background}\label{sec:math_background}

In this section, we review the concepts of two key components in our proposed FDA-QC method, namely the functional shape data analysis (FDA)~\cite{srivastava2010shape,srivastava2016functional} and quasi-conformal (QC) theory~\cite{lehto1973quasiconformal,ahlfors2006lectures}.

\begin{figure}[t]
    \centering
    \includegraphics[width=\linewidth]{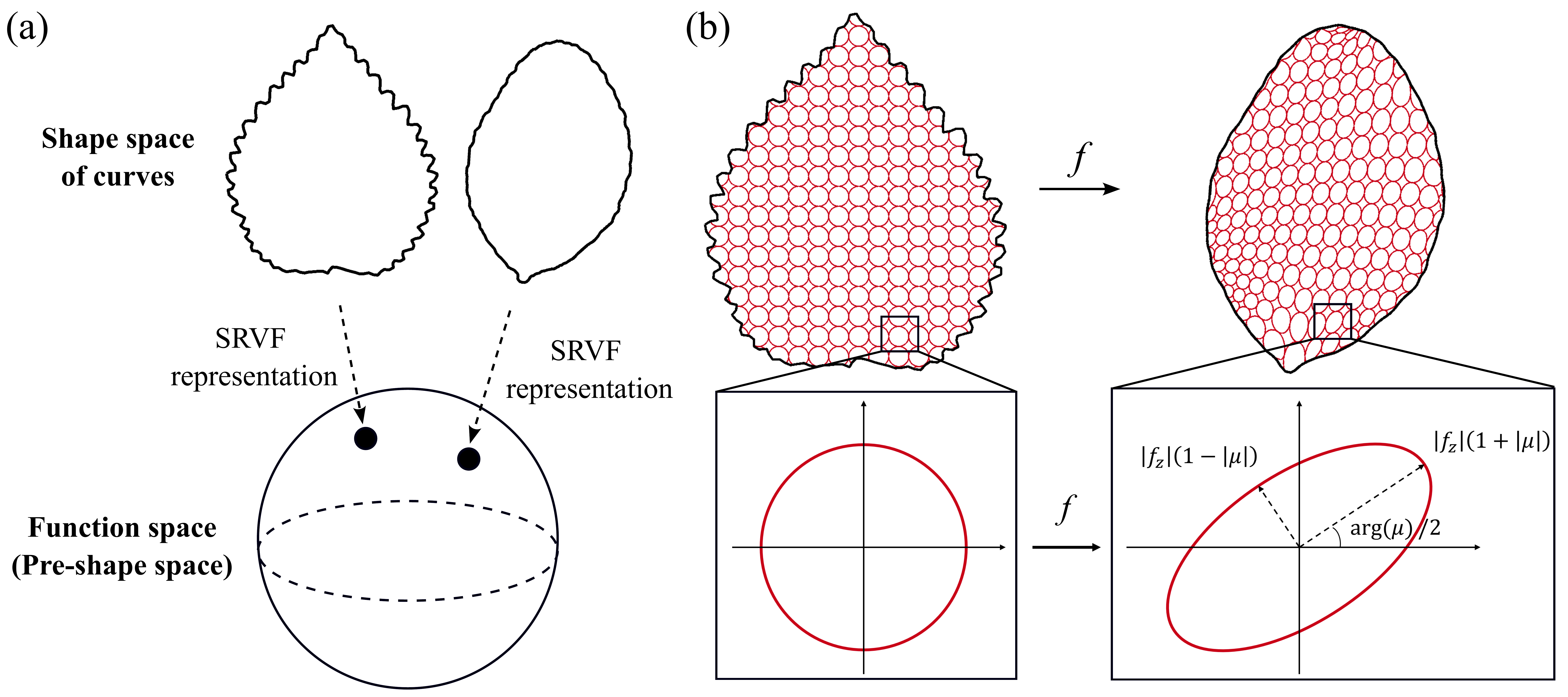}
    \caption{\textbf{An illustration of the basic concepts in functional shape data analysis and quasi-conformal theory.} (a) Two curves in the Euclidean space can be represented as the square-root velocity functions (SRVF) in the function space (also known as the pre-shape space), in which the optimal registration between them can be effectively obtained. (b) A quasi-conformal map $f$ between two planar structures can be visualized as the deformation of small circles to small ellipses. The maximal magnification, maximal shrinkage, and rotation of the small ellipses can be expressed in terms of the Beltrami coefficient $\mu$ associated with the quasi-conformal map, which encodes the overall quasi-conformal distortion of $f$ and hence the shape difference between the two planar structures.}
    \label{fig:concept_fda_qc}
\end{figure}

\subsection{Functional shape data analysis (FDA)}
Let $\gamma:[0,1]\to\mathbb R^2$ be a regular parameterized planar curve with $\|\dot\gamma(t)\|>0$. The square-root velocity function (SRVF) representation of $\gamma$ is defined by
\begin{equation} \label{eqt:srvf}
q(t)=\frac{\dot\gamma(t)}{\sqrt{\|\dot\gamma(t)\|}}.
\end{equation}
Conversely, given a SRVF $q(t)$, one can reconstruct the corresponding curve $\gamma(t)$ by
\begin{equation}
\gamma(t)=\gamma(0)+\int_0^t q(s)\|q(s)\|\,ds.
\end{equation}
In other words, $\gamma$ can be uniquely determined by the SRVF up to a translation.

Also, if $\gamma$ is a closed curve, we have $\gamma(0) = \gamma(1)$. Correspondingly, we have the closure condition for the SRVF: 
\begin{equation} \label{eqt:closure}
    \int_0^1 q(s)\|q(s)\|\,ds = 0.
\end{equation}

Note that the SRVF enables us to represent curves as mathematical functions and study them in the function space (see Fig.~\ref{fig:concept_fda_qc}(a)). In particular, after removing translation and normalizing the curve length, shape comparison between curves can be carried out via optimal rotations and orientation-preserving reparameterizations in the SRVF representation space (also known as the pre-shape space).

More specifically, given two curves $\gamma_{\mathcal M}$ and $\gamma_{\mathcal N}$, one can first rescale them to be with unit length. Then, let $\tilde q_{\mathcal M}$ and $\tilde q_{\mathcal N}$ denote their unit-norm SRVFs. The optimal elastic registration between them can be obtained by solving
\begin{equation} \label{eqt:optimalregistration}
(R^\ast,\varphi^\ast)=\arg\max_{R,\varphi}\;
\Big\langle \tilde q_{\mathcal M},\;R\big((\tilde q_{\mathcal N}\circ\varphi)\sqrt{\dot\varphi}\big)\Big\rangle,
\end{equation}
where $R\in\mathrm{SO}(2)$ is a rotation and $\varphi\in\mathrm{Diff}^+(\mathbb S^1)$ is an orientation-preserving reparameterization. 

In practice, the above registration problem can be solved using dynamic programming~\cite{srivastava2016functional}. The resulting optimal pair $(R^\ast,\varphi^\ast)$ induces a monotone, orientation-preserving correspondence $g:\gamma_{\mathcal M} \to \gamma_{\mathcal N}$, which provides a homeomorphic matching between the two curves.

\subsection{Quasi-conformal (QC) mappings}

Mathematically, a quasi-conformal map $f$ between two planar domains satisfies the Beltrami equation 
\begin{equation} \label{eqt:beltramieqt}
    \frac{\partial f}{\partial \overline{z}} = \mu(z) \frac{\partial f}{\partial z}
\end{equation}
for some complex-valued function $\mu$ with $\|\mu\|_{\infty} < 1$. Here, the Wirtinger derivatives $\frac{\partial f}{\partial \overline{z}}$ and $\frac{\partial f}{\partial z}$ are given by
\begin{equation}
     \frac{\partial f}{\partial \overline{z}} = f_{\overline{z}}= \frac{1}{2}\left(\frac{\partial f}{\partial x} + i \frac{\partial f}{\partial y}\right)
\end{equation} 
and
\begin{equation}
      \frac{\partial f}{\partial z} = f_{z} = \frac{1}{2}\left(\frac{\partial f}{\partial x} - i\frac{\partial f}{\partial y}\right),
\end{equation} 
where $z = x+iy$.

Note that the complex-valued function $\mu(z)$ in Eq.~\eqref{eqt:beltramieqt} is called the Beltrami coefficient of $f$, and it encodes key geometric information of the mapping $f$. 

Specifically, if we consider the first-order approximation of $f$ around a point $z_0$, we have
\begin{equation}
\begin{split}
    f(z) &\approx f(z_0) + f_z(z_0) (z-z_0) + f_{\overline{z}}(z_0) \overline{z-z_0}\\
    & = f(z_0) + f_z(z_0)\left(z-z_0 + \mu(z_0) \overline{z-z_0}\right),
\end{split}
\end{equation}
and hence
\begin{equation}
    |f(z)-f(z_0)| \approx |f_z(z_0)| (1+|\mu(z_0)|) |z-z_0|.
\end{equation}
Consequently, under the mapping $f$, an infinitesimal circle centered at $z_0$ will be mapped to an infinitesimal ellipse centered at $f(z_0)$ (see Fig.~\ref{fig:concept_fda_qc}(b) for an illustration), with the maximum magnification given by $|f_z(z_0)|(1+|\mu(z_0)|)$, the maximum shrinkage given by $|f_z(z_0)|(1-|\mu(z_0)|)$, and the pointwise quasi-conformal  dilatation given by
\begin{equation}
 K(z_0) =\frac{1+|\mu(z_0)|}{1-|\mu(z_0)|}.
\end{equation}

Besides, given certain boundary constraints and a Beltrami coefficient $\mu$, we can compute a quasi-conformal map associated with $\mu$ by solving a set of sparse linear systems via the Linear Beltrami Solver (LBS) method~\cite{lam2014landmark}. More specifically, we can obtain a quasi-conformal map $f = u + iv$ associated with the Beltrami coefficient $\mu$ by solving the following generalized Laplace equations subject to the given boundary constraints:
\begin{equation} \label{eqt:lbs}
\left\{\begin{array}{l}
\nabla \cdot(A_{\mu} \nabla u)=0, \\
\nabla \cdot(A_{\mu} \nabla v)=0,
\end{array}\right.
\end{equation}
where 
$
A_{\mu}=\left(\begin{array}{cc}
\alpha_1 & \alpha_2 \\
\alpha_2 & \alpha_3
\end{array}\right)
$
with 
\begin{equation}
    \alpha_1=\frac{(\text{Re}({\mu})-1)^2+(\text{Im}({\mu}))^2}{1-|\mu|^2},  \ \ \alpha_2=-\frac{2 \ \text{Im}({\mu})}{1-|\mu|^2},  \ \ \alpha_3=\frac{(\text{Re}({\mu})+1)^2+(\text{Im}({\mu}))^2}{1-|\mu|^2}.
\end{equation}
In the discrete case, the planar domains are represented as triangle meshes. The elliptic PDEs in Eq.~\eqref{eqt:lbs} become a sparse positive
definite linear system and can be efficiently solved.

\section{Proposed Methods} \label{sect:main}

In this section, we describe our proposed methods. We start by introducing our FDA-QC mapping method that incorporates functional data analysis and quasi-conformal theory. We then describe how the proposed FDA-QC mapping method can be naturally utilized for achieving continuous shape deformations between planar shapes. Finally, we describe how the FDA-QC mapping method can be used for establishing a shape dissimilarity measure for planar objects for shape analysis.

\subsection{The FDA-QC mapping method}\label{sec:shape_mapping}

Let $\mathcal{M}, \mathcal{N}$ be two simply connected planar shapes with boundaries $\gamma_{\mathcal{M}}$ and $\gamma_{\mathcal{N}}$ respectively. Our goal is to construct a planar mapping $f:\mathcal{M} \to \mathcal{N}$ that optimizes both the boundary correspondence $\gamma_{\mathcal{M}} \leftrightarrow \gamma_{\mathcal{N}}$ and the interior shape correspondence $\text{int}(\mathcal{M}) \leftrightarrow \text{int}(\mathcal{N})$. Optionally, interior landmarks $\{p^{\mathcal{M}}_i\}_{i=1}^n$ and $\{p^{\mathcal{N}}_i\}_{i=1}^n$ representing common prominent features on the two shapes can also be incorporated into our mapping method.

\subsubsection{Boundary mapping using the SRVF representation}
We first establish a correspondence between the boundary curves $\gamma_{\mathcal M}$ and $\gamma_{\mathcal N}$. As introduced in Section~\ref{sec:math_background}, we represent the two boundary curves using the SRVF formulation and compute their elastic registration, yielding an optimal rotation $R^\ast$ and reparameterization $\varphi^\ast$.

More specifically, we first remove translation and normalize the boundary lengths of $\gamma_{\mathcal M}$ and $\gamma_{\mathcal N}$, so that the corresponding SRVFs $\tilde q_{\mathcal M}$ and $\tilde q_{\mathcal N}$ obtained from Eq.~\eqref{eqt:srvf} are of unit norm. We then compute the optimal alignment by solving Eq.~\eqref{eqt:optimalregistration}, with the closure condition in Eq.~\eqref{eqt:closure} further enforced by a projection step. The resulting optimal SRVF registration induces the boundary correspondence $g:\partial\mathcal M\to\partial\mathcal N$. This map is monotone and orientation-preserving, and hence provides a homeomorphic matching between the two boundaries.

Note that the optimality of the curve registration obtained via the SRVF representation has a rigorous theoretical basis. Specifically, the SRVF transform converts the elastic metric on parameterized curves into the standard $L^2$ metric, so that the search for the best alignment under rotations and reparameterizations becomes a well-defined variational problem on shape space~\cite{srivastava2010shape,srivastava2016functional}. Hence, the resulting correspondence is geometrically meaningful and provides a principled boundary condition for the subsequent quasi-conformal interior mapping.

In the discrete case, the curves and hence the corresponding SRVFs are represented by discrete points. Fully solving Eq.~\eqref{eqt:optimalregistration} requires considering all possible shifts of the vertex indices of the SRVFs, followed by finding an optimal rotation via a singular value decomposition and finding an optimal reparameterization via dynamic programming for each index shift, which is computationally expensive. To simplify the process, we consider first finding the optimal index shift and the optimal rotation without solving the dynamic programming problem. Then, fixing the optimal index shift and the optimal rotation in the SRVF space, we solve for the optimal reparameterization via dynamic programming. The result obtained using this simplified procedure is used as the baseline. We then run the more complete optimization procedure involving index shifts, rotation, and reparameterization simultaneously on a sparse subset of possible shifts. The best alignment result among the baseline and all runs is taken as the final registration result. This simplified process allows us to accelerate the computation without largely affecting the optimality of the registration.

\subsubsection{Interior mapping using quasi-conformal map}
After finding the optimal correspondence between the boundary curves, we proceed to the problem of establishing the correspondence between the entire planar domains $\mathcal{M}$ and $\mathcal{N}$.

While many quasi-conformal mapping methods have been developed in recent years, they usually require a prescribed boundary correspondence in the computation of the mapping between the domains. For instance, the method in~\cite{choi2020tooth} computes quasi-conformal maps between planar rectangular domains with the correspondences between the corners fixed. In~\cite{choi2018planar}, the mappings between planar insect wing shapes also rely on some prescribed boundary landmark correspondences based on the vein structure of the insect wings. For more general planar shapes, it may be difficult to identify such boundary correspondences manually. Nevertheless, this technical challenge can be resolved by utilizing the curve registration result obtained via the above-mentioned boundary mapping.

Here, we propose to use the optimal elastic curve registration $g:\gamma_{\mathcal{M}} \to \gamma_{\mathcal{N}}$ that we obtained above using FDA as the boundary correspondence for the mapping problem between $\mathcal{M}$ and $\mathcal{N}$. More specifically, we compute a quasi-conformal map $f:\mathcal{M} \to \mathcal{N}$ by solving Eq.~\eqref{eqt:lbs} subject to the FDA-based boundary condition
\begin{equation}
    f|_{\partial{\mathcal{M}}} = g.
\end{equation}
In case there are also prescribed interior landmarks $\{p^{\mathcal{M}}_i\}_{i=1}^n$ and $\{p^{\mathcal{N}}_i\}_{i=1}^n$ on the two shapes, we further enforce the pointwise constraint
\begin{equation}
    f(p^{\mathcal{M}}_i) = p^{\mathcal{N}}_i,
\end{equation}
for all $i = 1, 2, \dots, n$.

Now, note that in the formulation in Eq.~\eqref{eqt:lbs}, the matrix $A_{\mu}$ depends on a given Beltrami coefficient $\mu$. To seek a mapping between $\mathcal{M}$ and $\mathcal{N}$ with the geometric distortion as low as possible, it is natural to consider the Beltrami coefficient $\mu = 0$, which is associated with a conformal map. Specifically, we have $
A_{0}=\left(\begin{array}{cc}
1 & 0 \\
0 & 1
\end{array}\right)
$ and hence Eq.~\eqref{eqt:lbs} reduces to the standard Laplace equation. However, this choice of $\mu = 0$ may not be compatible with the prescribed boundary constraints and interior landmark constraints. Therefore, directly solving Eq.~\eqref{eqt:lbs} with the given boundary constraints and interior landmark constraints may result in a non-bijective mapping containing local overlaps. We then locate these overlaps via the Beltrami coefficient of the resulting mapping and correct them via a truncate-and-reconstruct procedure. In particular, by computing the Beltrami coefficient $\mu^0$ of the resulting mapping $f$, we can locate the local overlaps at the points $z$ in the domain with $|\mu^0(z)| > 1$. To correct the overlaps, we consider a truncation operation on the Beltrami coefficient using the following truncation function $\mathbb{I}$: 
\begin{equation}\label{eqt:truncation}
\mathbb I(\mu^0(z))= \begin{cases} \mu^0(z) & \text { if } |\mu^0(z)|< 1, \\ \delta \frac{\mu^0(z)}{|\mu^0(z)|} & \text { if } |\mu^0(z)| \geq 1,\end{cases}
\end{equation}
where $\delta$ is a truncation parameter. In practice, we set $\delta = 0.9$. 

Also, note that the truncation operation may lead to discontinuity in the Beltrami coefficient field. To alleviate this issue, we follow the idea of the Teichm\"uller mapping method~\cite{lui2014teichmuller,choi2018planar} and construct a smoothed Beltrami coefficient $\widetilde{\mu}$ by replacing the magnitude of $\mathbb I(\mu^0(z))$ at every $z$ with the average magnitude over the entire domain. Using the smoothed Beltrami coefficient $\widetilde{\mu}$, we solve Eq.~\eqref{eqt:lbs} again together with the FDA-based boundary constraints and interior landmark constraints. We then repeat the above truncation, smoothing, and mapping computation procedure until all local overlaps are resolved. Note that the theoretical convergence of this iterative procedure follows from~\cite{lui2015convergence}. However, in the discrete case, resolving the overlaps near the fixed landmark points requires the flexibility of adjusting the positions of sufficiently many vertices nearby, which is practically limited by the input mesh resolution. Therefore, if the mesh overlaps are not resolved after a maximum number of iterations (set to 100), we relax the requirement and construct a quasi-conformal map with the latest truncated Beltrami coefficient without further enforcing the landmark constraints. In this case, since the truncated Beltrami coefficient is obtained from the last landmark-matching mapping, this construction effectively yields a bijective, soft landmark-matching mapping result. This completes the proposed FDA-QC mapping method.

Altogether, we obtain an optimal registration map $f:\mathcal{M} \to \mathcal{N}$ between the two planar domains that takes both the optimality of the boundary correspondence and the interior shape deformation into consideration via the SRVF and Beltrami coefficient geometric representations.

\subsection{Shape morphing via FDA-QC} \label{sec:shape_morphing}

Note that the above-mentioned FDA-QC mapping method effectively produces a registration between two planar structures. Moreover, it offers a natural way to construct a continuous shape deformation between $\mathcal{M}$ and $\mathcal{N}$ in a fully automatic manner without requiring any other information at the intermediate states. In particular, instead of performing a linear interpolation between the vertices of the two structures, we make use of the FDA and QC theories for achieving the desired morphing. The key idea is to utilize the geodesics between the SRVFs in the function space for the boundary morphing and an interpolation of the Beltrami coefficient field for the interior morphing between the two shapes.

\subsubsection{Boundary evolution via SRVF geodesics}
Specifically, we first note that the geodesic path between the two SRVFs $q_{\mathcal{M}}$ and $q_{\mathcal{N}}$ in the function space is given by $\{ q_{\tau}\}_{\tau \in [0,1]}$ as follows~\cite{srivastava2016functional}:
\begin{equation}\label{eqt:fdamorphing}
    q_{\tau} = \frac{1}{\sin \theta}\left(\sin ( (1-\tau)\theta) q_{\mathcal{M}} + \sin (\tau \theta) q_{\mathcal{N}}\right),
\end{equation}
where $\theta = \cos^{-1} (\langle q_{\mathcal{M}}, q_{\mathcal{N}} \rangle)$. 

Now, for the two given planar structures $\mathcal{M}$ and $\mathcal{N}$, we first consider their boundaries $\partial \mathcal{M}$ and $\partial \mathcal{N}$. Using the unit-norm SRVF $\tilde q_{\mathcal M}$ and the elastic registration $R^\ast\big((\tilde q_{\mathcal N}\circ\varphi^\ast)\sqrt{\dot\varphi^\ast}\big)$ from Eq.~\eqref{eqt:optimalregistration}, we can apply Eq.~\eqref{eqt:fdamorphing} and obtain $q_{\tau}$ for any $\tau \in [0,1]$. Consequently, we can reconstruct a planar curve $\gamma_{\tau}$ as an intermediate curve in the continuous deformation between $\gamma_{\mathcal{M}}$ and $\gamma_{\mathcal{N}}$, with $\tau =0$ corresponding to $\gamma_{\mathcal{M}}$ and $\tau =1$ corresponding to $\gamma_{\mathcal{N}}$. This construction preserves the elastic registration principle, with the boundary features of the biological shapes moving along a geodesically consistent trajectory rather than by pointwise linear interpolation.

\subsubsection{Interior evolution via Beltrami coefficient field interpolation}
Now, for the deformation of the overall shape including the interior part, we construct a family of quasi-conformal mappings that are capable of transporting interior points consistently with the boundary shape changes. Recall that the quasi-conformal mapping $f:\mathcal{M} \to \mathcal{N}$ obtained from the FDA-QC mapping method is associated with a Beltrami coefficient $\mu$, which encodes the quasi-conformal distortion of $f$. To produce the intermediate shape deformations, we can consider dividing the emergence of the quasi-conformal shape changes in time. 

More specifically, we construct a family of intermediate QC maps $\{f_{\tau}\}_{\tau\in[0,1]}$ by scaling the quasi-conformal distortion field represented by the Beltrami coefficient:
\begin{equation}
\mu_{\tau}=\tau\,\mu.
\end{equation}
Intuitively, the overall quasi-conformal shape difference between $\mathcal{M}$ and $\mathcal{N}$ can be transformed into a continuous deformation process. For each $\tau$, we can reconstruct a QC map $f_{\tau} = u_{\tau} + i v_{\tau}$ with the associated Beltrami coefficient $\mu_{\tau}$ using the LBS method as in Eq.~\eqref{eqt:lbs}:
\begin{equation} \label{eqt:lbs_tau}
\left\{\begin{array}{l}
\nabla \cdot(A_{\mu_{\tau}} \nabla u_{\tau})=0, \\
\nabla \cdot(A_{\mu_{\tau}} \nabla v_{\tau})=0,
\end{array}\right.
\end{equation}
subject to the boundary constraint 
\begin{equation}
    f_{\tau}|_{\partial\mathcal M}=\gamma_{\tau}.
\end{equation}
It is easy to check that for $\tau =0$, we have $\tau\mu = 0$ and $\gamma_0 = \gamma_{\mathcal{M}}$, and hence 
\begin{equation}
    f_0(\mathcal{M}) = \mathcal{M}.
\end{equation}
In other words, $f_0$ is an identity map. Also, for $\tau = 1$, we have $\tau \mu = \mu$ and $\gamma_1 = \gamma_{\mathcal{N}}$, and hence 
\begin{equation}
    f_1(\mathcal{M}) = \mathcal{N} = f(\mathcal{M}).
\end{equation}
For $\tau \in (0,1)$, the intermediate shapes represented as $\mathcal M_{\tau}=f_{\tau}(\mathcal M)$ give a continuous morphing that effectively couples boundary registration with interior deformation. We remark that with the presence of landmark constraints in the computation of the FDA-QC mapping $f:\mathcal{M} \to \mathcal{N}$, an interpolation of the Beltrami coefficient field will automatically provide a smooth trajectory between every pair of landmarks in the intermediate mappings. Hence, the landmark constraints are not required in solving Eq.~\eqref{eqt:lbs_tau}.

Altogether, the FDA-QC mapping approach not only provides us with the registration between two shapes but also a natural way to infer the intermediate changes between two shapes and model the continuous growth process of planar biological structures via the shape space encoded by the SRVFs and Beltrami coefficients.

\subsection{Shape dissimilarity measure and shape analysis via FDA-QC}\label{sec:shape_analysis}

After establishing the registration and continuous morphing between planar domains by combining FDA and QC, we proceed to utilize FDA-QC for measuring the shape dissimilarity between planar biological structures and performing clustering analysis to understand their geometric variations.

\subsubsection{Dissimilarity measure induced by FDA-QC}
Note that by the theory of functional shape data analysis, there is an equivalence between a metric in the spaces of the SRVF representations and a metric in the shape space of the curves~\cite{srivastava2016functional}. Therefore, for any two given planar shapes $\mathcal{M}, \mathcal{N}$, we can easily quantify the difference between their boundaries $\partial \mathcal{M} = \gamma_{\mathcal{M}}$ and $\partial \mathcal{N} = \gamma_{\mathcal{N}}$ using their corresponding unit-norm SRVFs $\tilde q_{\mathcal M}$ and $\tilde q_{\mathcal N}$. Specifically, after finding the optimal rotation and reparameterization $(R^\ast,\varphi^\ast)$ by solving Eq.~\eqref{eqt:optimalregistration}, the corresponding elastic geodesic distance between the boundary curves of $\mathcal{M}$ and $\mathcal{N}$ is given by
\begin{equation}\label{eqt:dfda}
d_{\mathrm{FDA}}(\mathcal{M}, \mathcal{N})
=\arccos\Big\langle \tilde q_{\mathcal M},\;R^\ast\big((\tilde q_{\mathcal N}\circ\varphi^\ast)\sqrt{\dot\varphi^\ast}\big)\Big\rangle.
\end{equation}

As for the quantification of the interior shape variation, note that by quasi-conformal Teichm\"uller theory~\cite{gardiner2000quasiconformal}, the shape difference between two planar domains $\mathcal{M}$ and $\mathcal{N}$ can be quantified using the Teichm\"uller metric 
\begin{equation}
    d_{\text{Teichm\"uller}}(\mathcal{M},\mathcal{N}) =  \frac{1}{2} \log K(f),
\end{equation}
where $K = \frac{1+\|\mu\|_{\infty}}{1-\|\mu\|_{\infty}}$ is the maximal dilatation of the mapping $f:\mathcal{M} \to \mathcal{N}$. Specifically, smaller values of $K$ (and hence smaller values of $d_{\text{Teichm\"uller}}$) correspond to a close-to-conformal deformation and indicate a lower level of local stretching in the mapping between the two shapes, while larger values of $K$ and $d_{\text{Teichm\"uller}}$ indicate stronger local anisotropic distortion in the mapping result, thereby effectively capturing the local geometric shape difference between the two shapes. However, note that in practice, directly using the supreme norm of the Beltrami coefficient $\mu$ of the FDA-QC mapping may lead to a biased result caused by some extreme deformations at a small localized region. To better represent the overall shape variation induced by our FDA-QC mapping, here we consider replacing $\|\mu\|_{\infty}$ with the average value of $|\mu|$ over the entire domain, i.e., $\text{mean}(|\mu|)$. Moreover, it is easy to see that for any $z \geq 0$, we have
\begin{equation}
    \log (z) = 2 \left(\frac{z-1}{z+1} + \frac{1}{3} \left(\frac{z-1}{z+1}\right)^3 + \frac{1}{5} \left(\frac{z-1}{z+1}\right)^5 + \cdots\right).
\end{equation}
Putting $z = \frac{1+|\mu|}{1-|\mu|}$ with $0 \leq |\mu|<1$, we have 
\begin{equation}
    \frac{z-1}{z+1} =  \frac{\frac{1+|\mu|}{1-|\mu|}-1}{\frac{1+|\mu|}{1-|\mu|}+1} = \frac{2|\mu|}{2} = |\mu|
\end{equation}
and hence
\begin{equation}
     \frac{1}{2} \log \left(\frac{1+|\mu|}{1-|\mu|}\right) = \frac{1}{2} \left(2 \left(|\mu| + \frac{1}{3}  |\mu|^3 +  \frac{1}{5} |\mu|^5 + \cdots\right)\right) \approx |\mu|.
\end{equation}
This motivates us to consider the following quasi-conformal shape dissimilarity measure for capturing the interior shape difference:
\begin{equation} \label{eqt:dqc}
    d_{\text{QC}} = \text{mean}(|\mu|).
\end{equation}

After establishing the above boundary-based and interior-based dissimilarity measures, it is natural for us to further consider the following combined measure:
\begin{equation} \label{eqt:d_combined}
    d_{\text{combined}} = \alpha d_{\text{FDA}} + \beta d_{\text{QC}},
\end{equation}
where $\alpha, \beta$ are two nonnegative weighting parameters. More specifically, note that by including the two dissimilarity measures $d_{\text{FDA}}$ and $d_{\text{QC}}$, the combined measure $d_{\text{combined}}$ can effectively capture both the boundary and interior shape variations. By changing the values of the weighting parameters $\alpha, \beta$, we can put more emphasis on either the boundary shape variation or the interior shape variation in the comparison between different planar biological and medical structures.

\subsubsection{Dissimilarity matrices and clustering analysis}
We can further extend the above formulation for quantifying the shape variation between a large group of shapes instead of only two shapes. More specifically, suppose we are given a set of $N$ planar biological shapes $\{\mathcal{M}_i\}_{i=1}^N$, possibly with additional feature correspondences encoded by $n$ landmark points $\{p^{\mathcal{M}_i}_k\}_{k=1}^{n}$ on each shape. For every pair of shapes \((\mathcal{M}_i,\mathcal{M}_j)\), where $i,j = 1, 2, \dots, N$ and $i \neq j$, we first compute a mapping $f_{ij}: \mathcal{M}_i \to \mathcal{M}_j$ using the FDA-QC mapping method (with the landmark constraints $f_{ij}(p^{\mathcal{M}_i}_k) = p^{\mathcal{M}_j}_k$ for $k = 1, 2, \dots, n$ also enforced if applicable). Then, for the boundary (FDA-based) dissimilarity, we have $d_{\text{FDA}}(\mathcal{M}_i,\mathcal{M}_j)$ from Eq.~\eqref{eqt:dfda}. 

As for the interior (QC-based) dissimilarity, we denote the Beltrami coefficient associated with the mapping $f_{ij}$ as $\mu_{ij}$. Following Eq.~\eqref{eqt:dqc}, the QC dissimilarity can be obtained by the average magnitude of the Beltrami coefficient $\mu_{ij}$. In particular, in the discrete case, $\mu_{ij}$ is discretized on the triangle elements. Since every triangle mesh in the dataset may have a different number of elements, we have 
\begin{equation}
\label{eq:qcdist_bee}
d_{\mathrm{QC}}(\mathcal{M}_i,\mathcal{M}_j)
=
\frac{1}{|\mathcal{F}_i|}
\sum_{T\in\mathcal{F}_i} |\mu_{ij}(T)|,
\end{equation}
where \(\mathcal{F}_i\) is the set of faces in the triangle mesh $\mathcal{M}_i$. Here, a lower value of \(d_{\mathrm{QC}}(\mathcal{M}_i,\mathcal{M}_j)\) indicates that $\mathcal{M}_i$ and $\mathcal{M}_j$ are related by a close-to-conformal deformation, while a larger value of \(d_{\mathrm{QC}}(\mathcal{M}_i,\mathcal{M}_j)\) indicate stronger local anisotropic distortion in the deformation.

From the FDA and QC dissimilarities between every pair of shapes, we obtain two $N\times N$ matrices $D^{\text{FDA}}$ and $D^{\text{QC}}$ with
\begin{equation}
    D_{ij}^{\text{FDA}} = d_{\text{FDA}}(\mathcal{M}_i,\mathcal{M}_j) \quad \text { and } \quad
    D_{ij}^{\text{QC}} = d_{\mathrm{QC}}(\mathcal{M}_i,\mathcal{M}_j).
\end{equation}
However, since the pairwise FDA and QC dissimilarities are computed in a directed manner via $f_{ij}$ and $f_{ji}$ and the resulting values may not be identical, we further symmetrize them by averaging the two directions. Specifically, we have the symmetrized matrices $\hat{D}^{\text{FDA}}$, $\hat{D}^{\text{QC}}$ with
\begin{equation}
    \hat{D}_{ij}^{\text{FDA}} = \frac{\hat{D}_{ij}^{\text{FDA}}+\hat{D}_{ji}^{\text{FDA}}}{2} \quad \text{ and } \quad \hat{D}_{ij}^{\text{QC}} = \frac{\hat{D}_{ij}^{\text{QC}}+\hat{D}_{ji}^{\text{QC}}}{2}. 
\end{equation}
Moreover, since the FDA and QC dissimilarities may have different scales, we normalize each of the two matrices $\hat{D}^{\text{FDA}}$, $\hat{D}^{\text{QC}}$ by a robust scale estimate (defined as the median of positive off-diagonal entries in each of them), yielding two symmetrized and normalized dissimilarity matrices \(\widetilde D^{\mathrm{FDA}}\) and \(\widetilde D^{\mathrm{QC}}\).

Following Eq.~\eqref{eqt:d_combined}, we can then define the $N\times N$ combined dissimilarity matrix \(D^{(\beta)}\) with the $(i,j)$-th entry $D_{ij}^{(\beta)}$ representing the fused dissimilarity between shape $i$ and shape $j$:
\begin{equation}
\label{eq:beta_distance}
D_{ij}^{(\beta)}
=
\beta\,\widetilde D^{\mathrm{FDA}}_{ij}
+
(1-\beta)\,\widetilde D^{\mathrm{QC}}_{ij},
\end{equation}
where $\beta\in[0,1]$. Here, \(\beta\) controls the contribution of the boundary term ($\widetilde D^{\mathrm{FDA}}_{ij}$), while \(1-\beta\) controls the contribution of the interior deformation term ($\widetilde D^{\mathrm{QC}}_{ij}$). In other words, the combined dissimilarity matrix reflects two distinct but related aspects of shape difference for the entire dataset. We remark that here we do not require metric axioms beyond symmetry and nonnegativity. Also, the choice of the parameter $\beta$ can be optimized, which can then effectively clarify the roles of the boundary and interior components in the shape variation. Specifically, the FDA part accounts for elastic variation of the boundaries of the shapes in the dataset modulo translation, scaling, rotation, and reparameterization, while the QC part describes the anisotropic interior deformation required by the induced correspondence. 

With the combined dissimilarity matrix \(D^{(\beta)}\), we can then easily perform clustering analysis by utilizing certain matrix conversion and community detection methods. To achieve this, we first convert \(D^{(\beta)}\) into a similarity matrix $W$ using a Gaussian kernel:
\begin{equation}
    W_{ij} = \exp\!\left( -\frac{D_{ij}^2}{2\sigma^2} \right),
\end{equation}
where the kernel width $\sigma$ is set to the median of the off‑diagonal distances in $D^{(\beta)}$. This makes the similarity decay adaptively to the scale of the values in \(D^{(\beta)}\). Also, we have $W_{ii}=1$ for all $i$ to represent self-similarities.

Next, we follow the approach in~\cite{choi2018planar} and apply adaptive thresholding to $W$ and perform community detection for clustering analysis. Specifically, the adaptive thresholding procedure aims to sparsify $W$ and retain only the most reliable connections between specimens in terms of their similarity. This is achieved by applying an adaptive, iterative row-wise thresholding procedure to $W$ as follows: For each row $i$, we first compute the row's threshold as
\begin{equation} \label{eqt:thresholding}
    \theta_i = \text{mean}(W_{i,:}) + \lambda \  \text{std}(W_{i,:}),
\end{equation}
where $\lambda$ is a thresholding parameter. Then, we can construct an $N\times N$ sparse matrix $S$ with 
\begin{equation} \label{eqt:sparseS}
    S_{ij} = \begin{cases}
        1 & \text{if } W_{ij} \geq \theta_i,\\
        0 & \text{otherwise.}
    \end{cases}
\end{equation}
We further symmetrize it by 
\begin{equation} \label{eqt:sparseS_sym}
    \widetilde{W} = \frac{S + S^T}{2}.
\end{equation} 
We can then replace $W$ with $\widetilde{W}$ and repeat Eq.~\eqref{eqt:thresholding}--\eqref{eqt:sparseS_sym} iteratively. It was proved in~\cite{choi2018planar} that the above adaptive thresholding procedure always converges in finitely many steps for any input similarity matrix $W$ and any thresholding parameter $\lambda$. In practice, we can set $\lambda = 1$ to keep only the sufficiently prominent similarity information and obtain the resulting thresholded matrix $\widetilde{W}$. Next, we apply the community detection method in~\cite{morrison2012discovering} and obtain the cluster labels for all specimens based on the thresholded matrix $\widetilde{W}$. Specifically, the method effectively accounts for the non-locality and asymmetry in the graph represented by $\widetilde{W}$ using the concept of the generalized Erd\H{o}s number~\cite{morrison2011asymmetric}, with the number of the resulting communities automatically determined. This completes the FDA-QC shape dissimilarity quantification and clustering analysis pipeline.

\section{Experiments} \label{sect:experiment}

In this section, we present several experiments to demonstrate the effectiveness of our proposed FDA-QC method. Specifically, we will showcase the use of FDA-QC for three tasks: shape mapping, shape morphing, and shape analysis. To demonstrate the wide applicability of our framework, we utilize a different biological dataset in each experiment presented below. The main parts of the proposed method are implemented in MATLAB, with a C implementation of the dynamic programming computation used via the MEX function. 

\subsection{Mapping between \emph{Acer palmatum} Japanese maple leaf shapes using FDA-QC}\label{subsec:mapping}

We first demonstrate the proposed FDA-QC mapping method on planar leaf shapes from an online leaf image dataset~\cite{leafuci}. Specifically, to compute a mapping between two leaf shapes, we first extract the leaf boundaries from the binary masks of the images and discretize them as closed polygonal curves. Then, we utilize the DistMesh tool~\cite{persson2004simple} to create discrete triangle meshes from the polygonal curves and denote the resulting meshes as $\mathcal{M}$ and $\mathcal{N}$. Then, as described in Section~\ref{sec:shape_mapping}, the FDA-QC method produces the resulting mapping using the FDA-based boundary registration step followed by the interior QC mapping step.

\begin{figure}[t!]
    \centering
    \includegraphics[width=\linewidth]{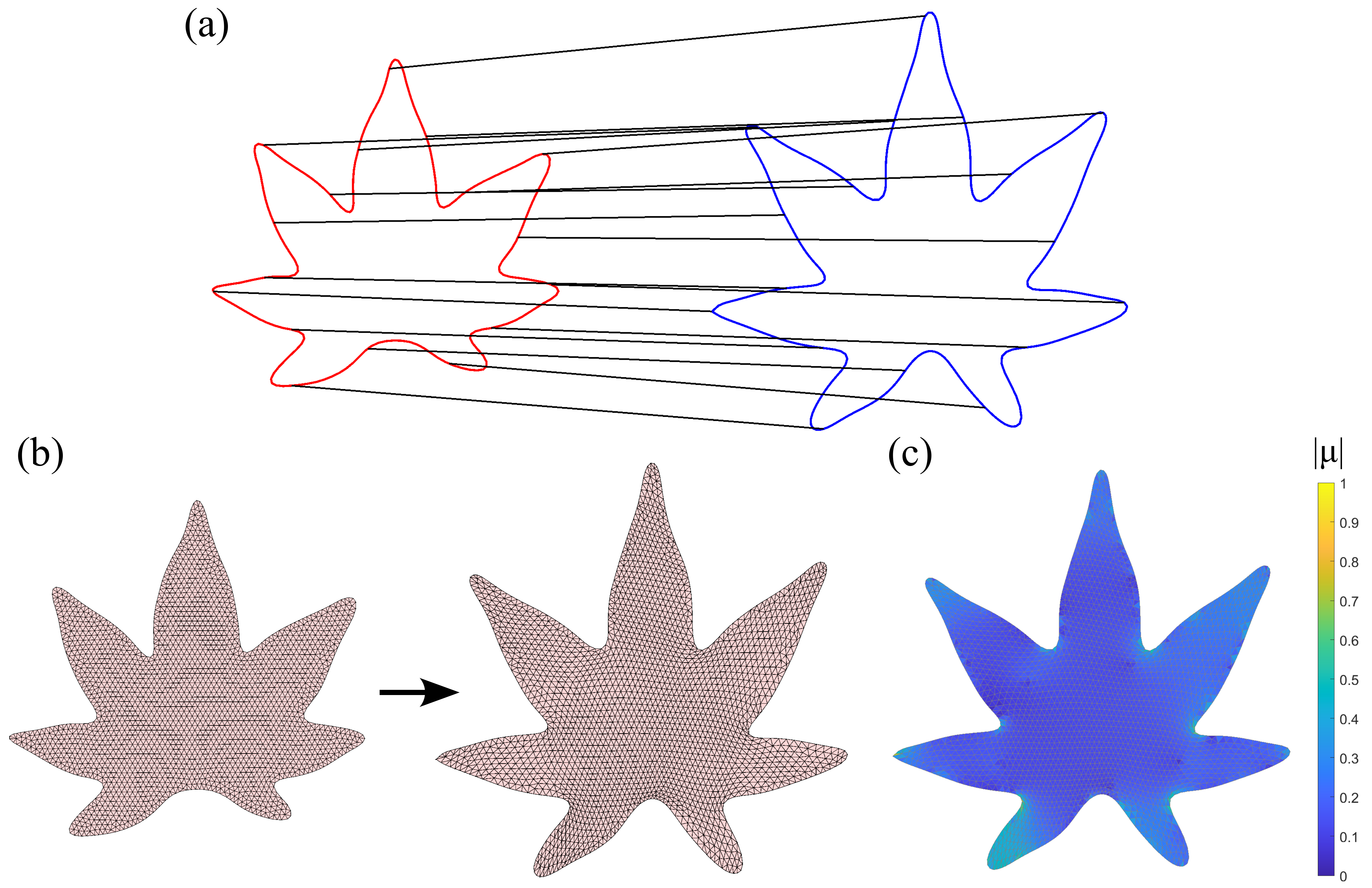}
    \caption{\textbf{Mapping a pair of \textit{Acer palmatum} Japanese maple leaf shapes using the proposed FDA-QC method.} (a)~FDA-based elastic boundary registration between a source leaf shape (red) and target leaf shape (blue). The black straight lines show the correspondence of several sample points established by the SRVF elastic registration. (b)~A triangle mesh of the source leaf shape and the FDA-QC mapping result of it onto the target shape. It can be observed that the overall shape is well matched, and the interior of the domain is deformed naturally. (c) A visualization of the Beltrami coefficient field of the FDA-QC mapping result. Here, all triangle elements are color-coded with the norm of the Beltrami coefficient $|\mu|$.}
    \label{fig:leaf_mapping}
\end{figure}

As shown earlier in Fig.~\ref{fig:illustration_fdaqc}, we can effectively compute a mapping between two leaves (a \emph{Urtica dioica} leaf and a \emph{Euonymus japonicus} leaf) with significantly different boundary shapes in a fully automatic manner. Here, we further present another example of mapping two \emph{Acer palmatum} (Japanese maple) leaf shapes, in which both the source and target shapes are highly non-convex. As shown in Fig.~\ref{fig:leaf_mapping}(a), using the elastic curve registration in the SRVF space, we can obtain an optimal boundary correspondence $g: \partial{\mathcal{M}} \to \partial{\mathcal{N}}$. In particular, it can be observed that the sharp tips of the leaves are automatically identified and well-matched. This shows the strength of the FDA-based boundary registration for revealing a smooth and biologically meaningful alignment of lobes and tips. 

We then use $g$ as the Dirichlet boundary condition in the QC mapping computation. As shown in Fig.~\ref{fig:leaf_mapping}(b), the triangle mesh constructed on the source shape can be effectively deformed to match the target shape. From the shape changes of the triangle elements, it can be observed that the overall shape deformation is smooth. From the plot of the norm of the Beltrami coefficient $|\mu|$ in Fig.~\ref{fig:leaf_mapping}(c), we can see that the quasi-conformal distortion is mainly concentrated near high-curvature boundary regions, while the interior exhibits only mild anisotropic stretching. 

Overall, this experiment confirms that combining FDA-based boundary registration with QC-based interior mapping yields high-quality planar correspondences that simultaneously respect detailed boundary features and preserve interior neighborhood structure. The FDA-QC mapping then forms the basis for the morphing and shape analysis experiments in the following subsections.

\subsection{Morphogenesis of \emph{Arabidopsis} leaves via FDA-QC}
As described previously in Section~\ref{sec:shape_morphing}, beyond producing a single registration map between two shapes, the proposed FDA-QC method naturally induces a continuous deformation from a source shape to a target shape.

To test the proposed framework, here we consider the \emph{Arabidopsis} leaf image dataset~\cite{saric2025computer} consisting of images of \emph{Arabidopsis} plants taken at the top view at different developmental time points. In particular, we take two images of the same \emph{Arabidopsis} plant at two developmental time points, in which four leaves of the plant grow in both size and shape. By extracting the boundaries of the leaves and treating the overall silhouette as a closed curve, we can discretize the images as simply connected triangle meshes $\mathcal M$ and $\mathcal N$ using DistMesh~\cite{persson2004simple} to study the growth of different leaves as well as the change of their relative positions over time. 

\begin{figure}[t]
\centering
\includegraphics[width=\textwidth]{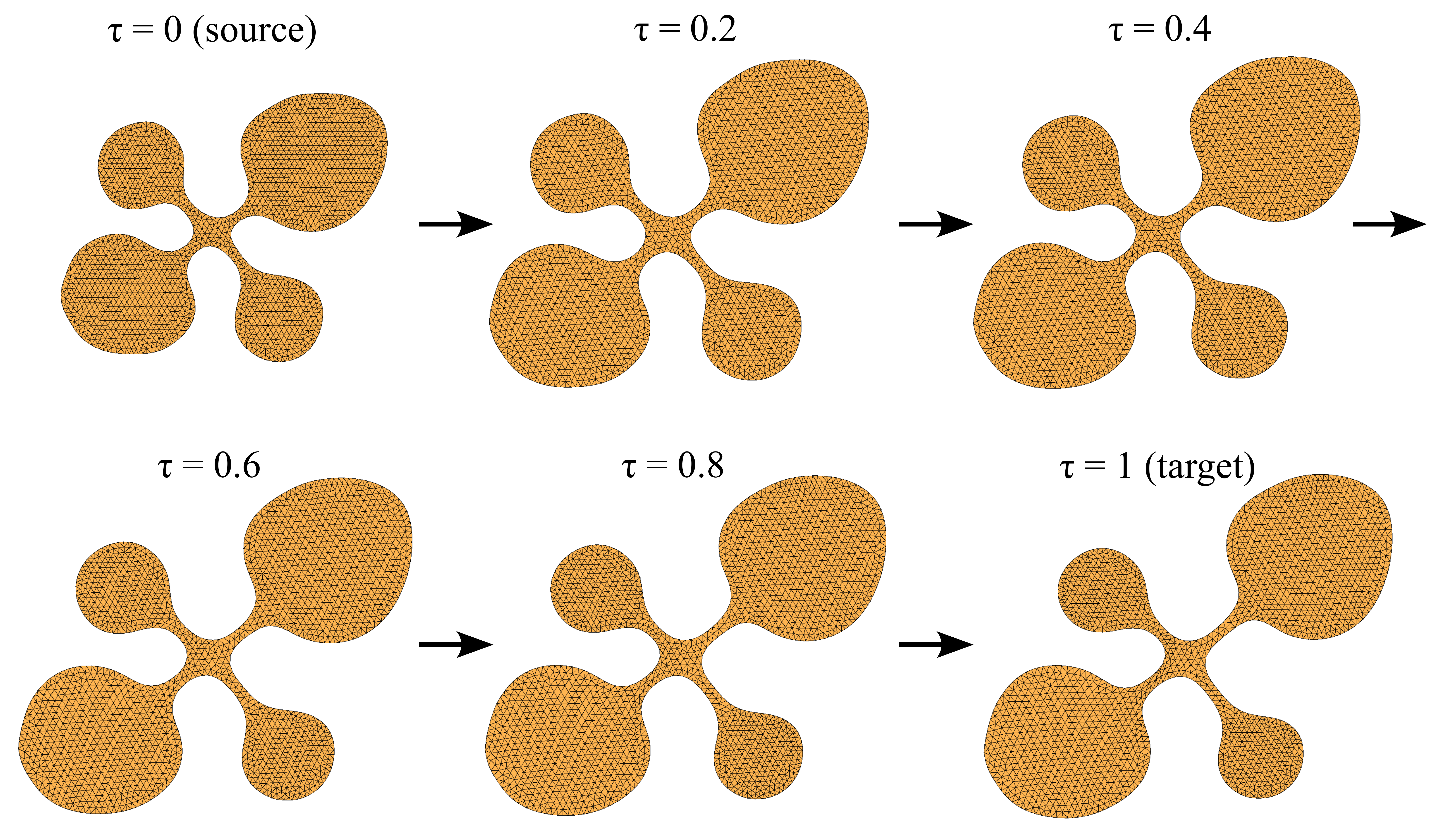}
\caption{\textbf{Shape morphing of \textit{Arabidopsis} leaves by the proposed FDA-QC framework.} Here, $\tau =0$ corresponds to the source leaf shape, $\tau = 0.2, 0.4, 0.6, 0.8, 1$ show the morphing results of it towards a target leaf shape. The boundary at each $\tau$ is obtained from the SRVF geodesic in the function space, while the interior deformation is generated by reconstructing a QC map from the interpolated Beltrami coefficient $\mu_{\tau}=\tau\mu$ with the boundary constraint $f_{\tau}|_{\partial\mathcal M}=\gamma_{\tau}$.}
\label{fig:leaf_morphing}
\end{figure}

To achieve this, we first compute a mapping between $\mathcal M$ and $\mathcal N$ using our FDA-QC mapping method. Then, by following the shape morphing procedure as detailed in Section~\ref{sec:shape_morphing}, we can consider the growth process as a one-parameter family of intermediate shapes indexed by $\tau\in[0,1]$, where $\tau=0$ corresponds to the initial time point and $\tau=1$ corresponds to the final time point. Each intermediate shape is then obtained by the FDA geodesic path in Eq.~\eqref{eqt:fdamorphing} followed by the QC mapping computation in Eq.~\eqref{eqt:lbs_tau}. 
Fig.~\ref{fig:leaf_morphing} shows six snapshots of the morphing sequence at $\tau = 0$, $0.2$, $0.4$, $0.6$, $0.8$, $1$ generated by the framework. From the morphing sequence, one can see that the initial triangle mesh undergoes a smooth deformation, with the boundary features evolving continuously along the FDA geodesic. Also, the interior part of the triangle mesh undergoes controlled anisotropic stretching dictated by the interpolated Beltrami coefficient field.

\begin{figure}[t]
    \centering
    \includegraphics[width=\linewidth]{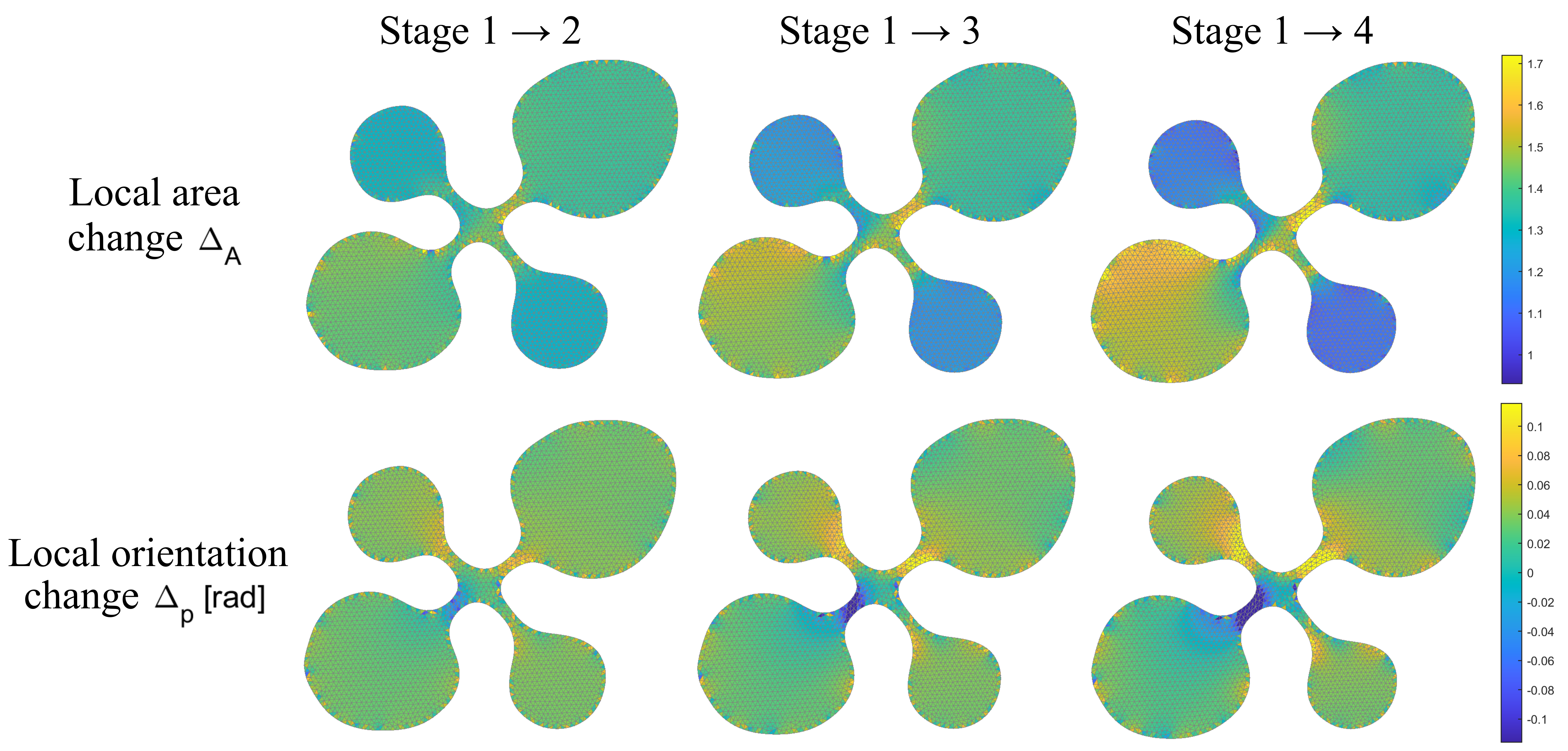}
    \caption{
    \textbf{Local area and orientation change during the temporal development of an \textit{Arabidopsis} plant captured by our FDA-QC morphing framework.} Each column corresponds to one temporal transition between different stages in Fig.~\ref{fig:leaf_morphing}. The top row shows the FDA-QC mapping result color-coded with the local area change \(\Delta_A\), while the bottom row shows the result color-coded with the local orientation change \(\Delta_p\) (in radian). The deformation fields are visualized on the target-stage mesh for each transition. For the \(\Delta_A\) plots, warm colors indicate stronger local expansion. For the \(\Delta_p\) plots, values close to zero indicate weak local rotation/reorientation.
    } 
    \label{fig:arabidopsis_temporal_qc}
\end{figure}

With this continuous geometric trajectory between two shapes, we can further analyze the local growth of the leaves at different developmental stages. In Fig.~\ref{fig:arabidopsis_temporal_qc} we extract the spatial intensity of areal expansion \(\Delta_A\) (given by the determinant of the Jacobian $J$ of the mapping $f_{\tau}$) and the local orientation change \(\Delta_p\) of the triangle elements (in radian, given by the polar decomposition of the Jacobian $J$) from the morphing results. It can be observed that at different developmental stages, the overall local area changes are different. One can also see that at each developmental stage, the area change is spatially heterogeneous. As for the local orientation change, the most prominent values are found at the central region of the mesh at the later developmental stage, indicating that the leaves are growing outward most significantly after a certain time point. 

Altogether, the proposed FDA-QC method offers a new way for understanding and modelling the local growth mechanisms underlying the biological image snapshots taken at different time points.

\subsection{Shape analysis of European honey bee (\emph{Apis mellifera}) wings}
\label{subsec:shape_analysis}

Next, we utilize our proposed FDA-QC framework for quantifying and analyzing the shape variation in a collection of biological shape data. Here, we apply the proposed method to the European honey bee (\emph{Apis mellifera}) wing dataset~\cite{oleksa2022collection,oleksa2023honey} to study whether region-level morphological variation is driven primarily by outer boundary geometry or by interior venation features. As described in Section~\ref{sec:shape_analysis}, our shape analysis pipeline quantifies the relative contribution of these two geometric components by combining pairwise geometric mappings and dissimilarity measures, optimization of the weighting parameters between the boundary-based measures and interior-based measures, and a clustering analysis.

\begin{figure}[t!]
    \centering
    \includegraphics[width=\linewidth]{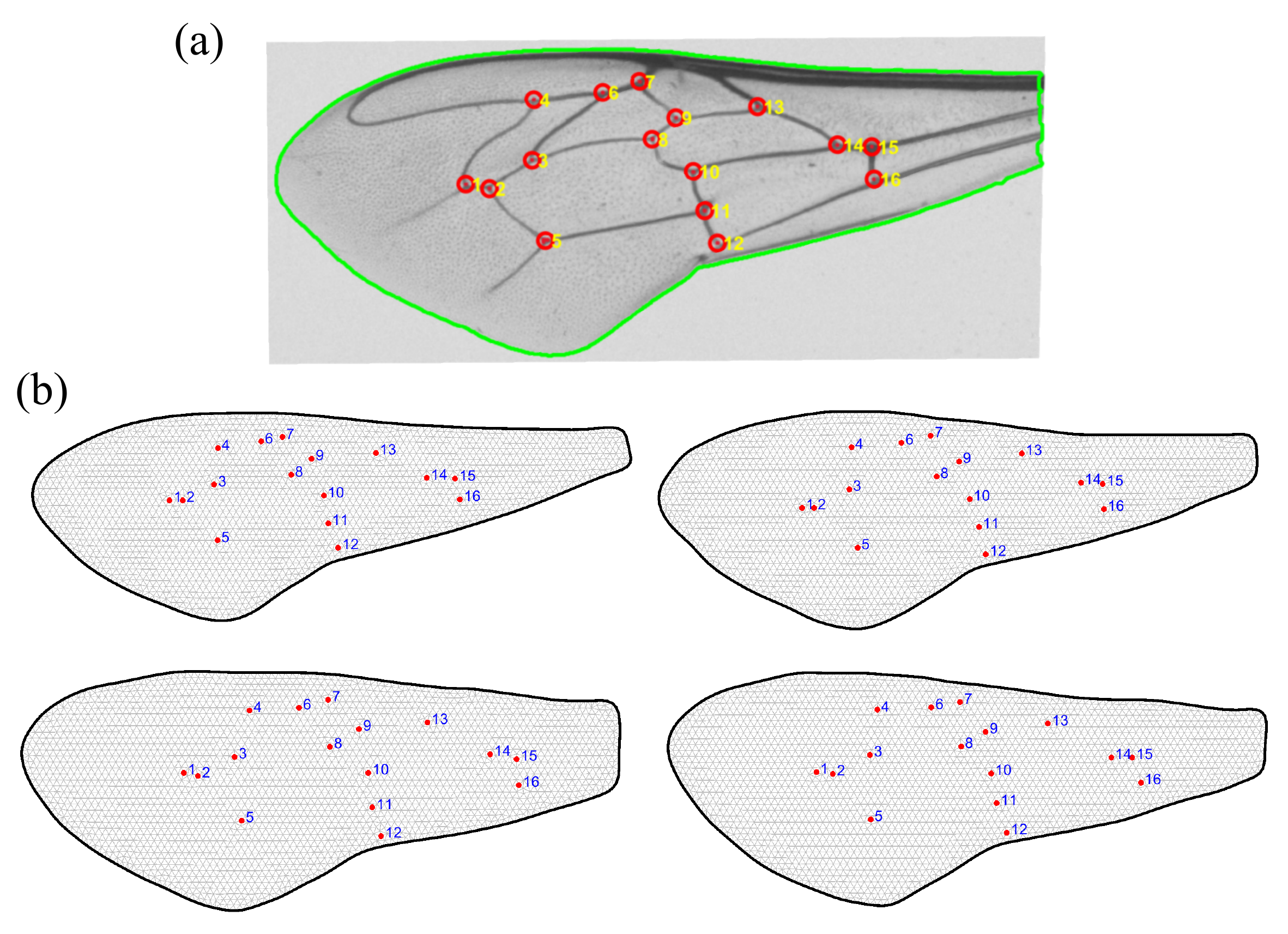}
    \caption{\textbf{Boundary and landmark extraction for honey bee (\textit{Apis mellifera}) wing shape analysis.} (a) Given a honey bee wing specimen, we extract the outer boundary curve (highlighted in green) and \(16\) stable venation branch/intersection points (highlighted in red). (b) Four examples of honey bee wing shapes discretized as triangle meshes. For each wing, the wing boundary is highlighted in black, and the 16 landmarks are highlighted in red and labelled.}
    \label{fig:honeybee_landmark}
\end{figure}

Specifically, we consider honey bee wing images from four geographic regions in the collection~\cite{oleksa2022collection,oleksa2023honey}: Austria (AT), Croatia (HR), Moldova (MD), and Slovenia (SI), with \(12\) specimens per region (total \(N=48\)). Each wing is modeled as a simply connected planar domain \(\mathcal{M}_i\), represented by an outer boundary \(\gamma_{\mathcal{M}_i}\) and a set of $n = 16$ stable venation branch/intersection points as landmarks \(P_{\mathcal{M}_i}=\{p^{\mathcal{M}_i}_k\}_{k=1}^{16}\) (see Fig.~\ref{fig:honeybee_landmark}(a) for an illustration). Note that the outer boundary effectively represents the overall shape of the honey bee wing, while the landmarks can be considered as interior geometric features encoding the relative organization of the venation network. After extracting the boundary contour and the interior landmarks from a honey bee wing image, we then generate a planar triangle mesh via DistMesh~\cite{persson2004simple}, with the labelled landmarks included as fixed vertices (see Fig.~\ref{fig:honeybee_landmark}(b) for examples). This yields a set of discrete triangle meshes for both the boundary registration and quasi-conformal mapping procedures. 

\begin{figure}[t]
    \centering
    \includegraphics[width=\linewidth]{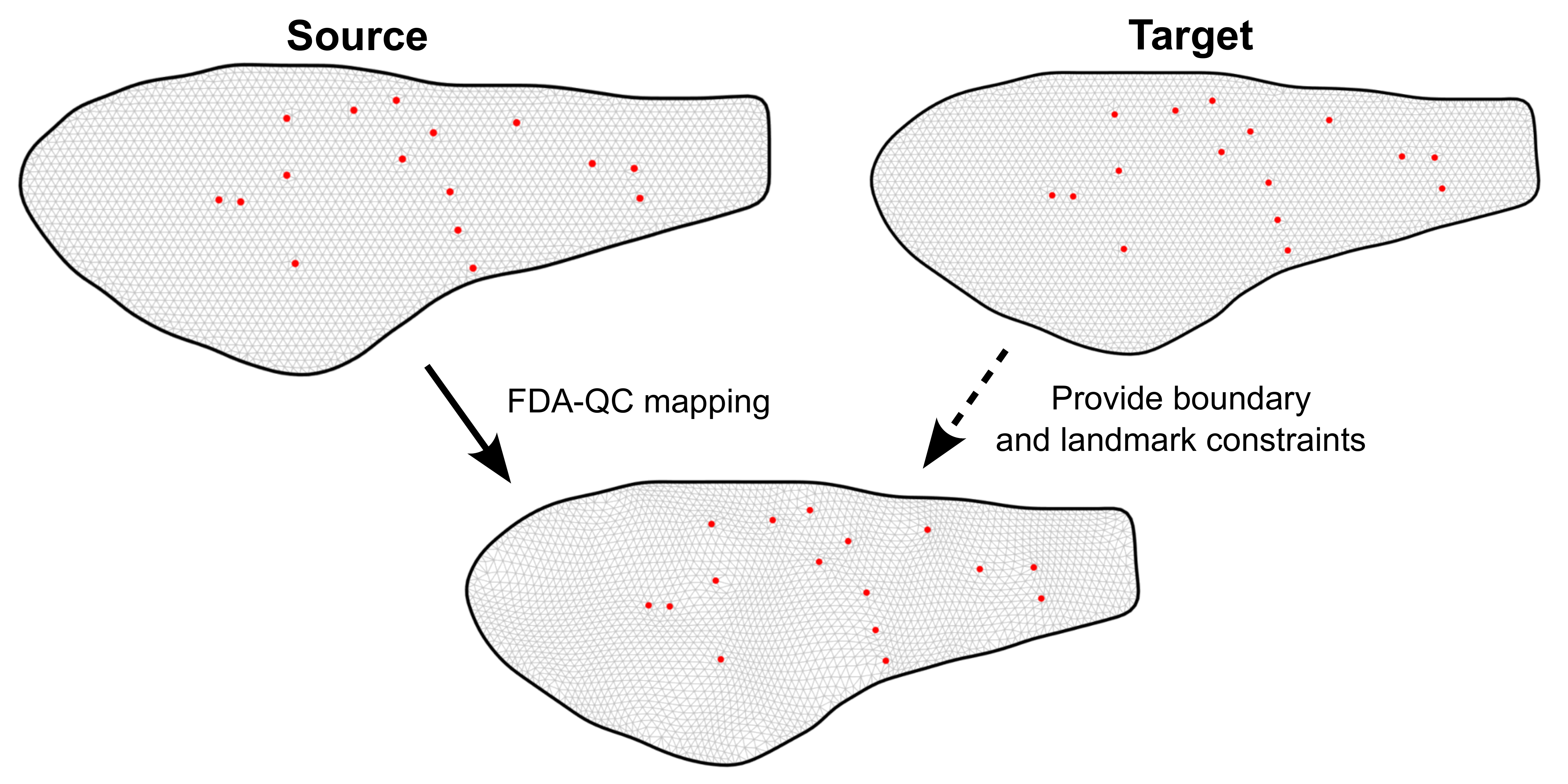}
    \caption{\textbf{FDA-QC workflow for honey bee wing (\textit{Apis mellifera}) shape analysis.} Given two wing shapes to be compared, we first extract the boundary and interior landmarks for each of them as described in Fig.~\ref{fig:honeybee_landmark}. Then, we construct a triangulation for each of them using DistMesh, resulting in the source and target meshes. By the pairwise FDA boundary registration and landmark-constrained QC mapping, we can obtain the FDA-QC mapping result from the source to the target, from which we can extract the shape dissimilarity measures for clustering analysis.}
    \label{fig:wing_mapping}
\end{figure}

Our goal is to use the FDA-QC framework to quantify shape variation among honey bee wings. In particular, we aim to assess whether the dominant signal of the shape variation is carried by the global boundary shape, the internal venation geometry, or both. To achieve this, we first apply our FDA-QC mapping method to compute an optimal mapping between every pair of wing shapes $\mathcal{M}_i, \mathcal{M}_j$ with $i, j = 1, 2, \dots, 48$ and $i \neq j$. In particular, we consider landmark-constrained QC mapping subject to the FDA-based boundary constraints and the correspondence of all 16 interior landmarks. The process is illustrated in Fig.~\ref{fig:wing_mapping}. After computing all pairwise mappings, we can construct a $48 \times 48$ combined shape dissimilarity matrix \(D^{(\beta)}\) with each $D_{ij}^{(\beta)}$ given by Eq.~\eqref{eq:beta_distance}. Following the matrix conversion, adaptive thresholding, and community detection procedures in Section~\ref{sec:shape_analysis}, we can perform a clustering analysis on the honey bee wings. We further use multidimensional scaling (MDS) as a qualitative visualization tool for the geometric variation of the specimens induced by our FDA-QC method, yielding a two-dimensional embedding of the specimens. We remark that here the MDS is only used for providing a low-dimensional visual summary of the shape dissimilarity induced by our FDA-QC method and is not used as a quantitative classifier. 

Recall that $\beta \in [0,1]$ is a weighting parameter controlling the contribution of the boundary term and the interior term in the combined dissimilarity measure. To select the optimal weight for the clustering analysis, we sweep \(\beta\) over uniform values in $[0,1]$ with an increment of $0.01$, i.e., $\beta = 0$, $0.01$, $\dots$, $0.99$, $1$. For each value, we compare the resulting clusters with the known region labels using the Adjusted Rand Index (ARI), Normalized Mutual Information (NMI), and purity. Note that the region labels are used only for external evaluation and for selecting \(\beta\), and they are not used in the graph construction or clustering steps in each trial. The optimal weight \(\beta^\ast\) is selected by maximizing ARI, with NMI and purity used as tie-breakers. If there is still a tie, the smallest $\beta$ is chosen. Under this criterion, we obtain
\begin{equation}
\beta^\ast = 0.75.
\end{equation}
At \(\beta^\ast\), the clustering agreement with region labels is given by ARI \(=0.6738\), NMI \(=0.8121\), and purity \(=0.7500\). In Table~\ref{tab:beta_comparison}, we further compare the outcome of $\beta = 0$ (i.e., only based on the QC dissimilarity term in Eq.~\eqref{eq:beta_distance}), $\beta = 1$ (i.e., only based on the FDA dissimilarity term in Eq.~\eqref{eq:beta_distance}), and the optimal $\beta^*$ and two other values. It can be observed that $\beta^* = 0.75$ gives the best ARI, NMI, and purity, indicating that a combined representation is more informative than either a boundary-only or an interior-only measure for studying the honey bee wing shape variation. 

\begin{table}[t]
    \centering
    \begin{tabular}{l||l|l|l}
        $\beta$ & ARI & NMI & Purity \\ \hline
        0 (interior only) & 0.4428 & 0.6335 & 0.5000\\
        0.7 & 0.6275 & 0.7528 & 0.7292\\
        0.75 (optimal) & 0.6738 & 0.8121 & 0.7500\\
        0.8 (tie) & 0.6738 & 0.8121 & 0.7500\\
        1 (boundary only) & 0.4428 & 0.6335 & 0.5000
    \end{tabular}
    \caption{\textbf{Honey bee wing shape clustering result achieved by different choices of $\beta$ in the dissimilarity matrix formulation.} Here, $\beta = 0$ corresponds to the case where the clustering is purely based on the interior quasi-conformal dissimilarity, while $\beta = 1$ corresponds to the case where the clustering is purely based on the SRVF-based boundary distance. $\beta^* = 0.75$ is the optimal result. The results for two other values $0.7$ and $0.8$ are also displayed.}
    \label{tab:beta_comparison}
\end{table}

\begin{figure}[t]
    \centering
    \includegraphics[width=0.7\linewidth]{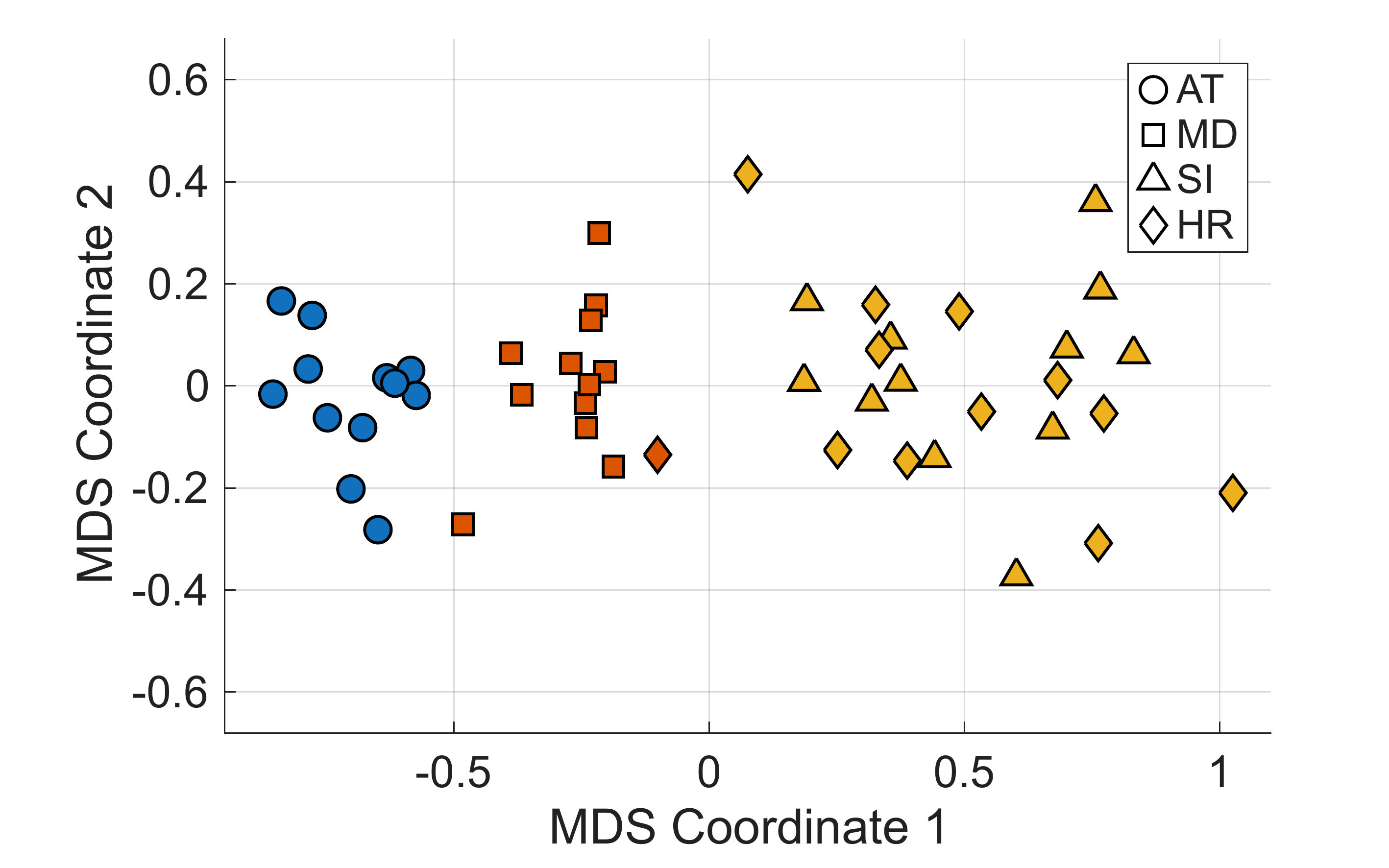}
    \caption{
    \textbf{MDS embedding of the \(48\) honey bee (\textit{Apis mellifera}) wing specimens using \(D^{(\beta^\ast)}\) with \(\beta^\ast=0.75\).} Marker shapes indicate geographic labels (AT, MD, SI, HR), and marker colors represent the clustering result obtained by the FDA-QC framework. The embedding shows compact AT and MD groups and an overlapping SI--HR sector, consistent with the European honey bee ecology. 
    }
    \label{fig:bee_mds}
\end{figure}

We can further study the cluster composition and the interpretation with respect to the four geographic groups. Note that the induced clustering structure is not uniform across regions. The geographic group AT is recovered as a single pure cluster (12/12), and MD is also recovered as a highly pure cluster (12/13), indicating relatively coherent shape organization under the fused dissimilarity measure. By contrast, SI and HR occupy an overlapping sector of the induced shape space ($12+11 = 23$ specimens), with one remaining HR specimen assigned to the MD-dominant cluster. The MDS embedding shown in Fig.~\ref{fig:bee_mds} is consistent with this observation: AT and MD occupy relatively compact regions in the MDS embedding, whereas SI and HR are less sharply separated. 

The above-mentioned clustering result can be explained from an ecological perspective. Specifically, note that Croatia (HR) and Slovenia (SI) are geographically adjacent and share several continuous ecological regions. As reported in prior studies~\cite{suvsnik2004molecular,munoz2009population,puskadija2021morphological}, the Croatian and Slovenian honey bee populations share clear morphological and genetic characteristics of the Carniolan honey bee, \textit{Apis mellifera carnica}. Therefore, the close grouping of Slovenian and Croatian samples in our analysis is biologically plausible. Note that while Austria (AT) is also geographically close to Slovenia, our clustering analysis shows a clear difference between the AT group and the HR--SI group. A possible explanation of their separation is their climate difference. Specifically, Austria is landlocked and largely dominated by the Alps, leading to a cooler climate, while Slovenia possesses an Alpine-Mediterranean climate. There are also differences in the management and breeding of honey bees~\cite{bouga2011review,brodschneider2019comparison}. As prior studies have shown that honey bee wing venation can vary with climate, geography, and beekeeping practices~\cite {ruttner1988biogeography,kaur2024influence,bakhchou2025morphological}, the shape variation between the AT wings and the SI wings observed in our analysis may be explained by these factors.

\begin{figure}[t]
    \centering
    \includegraphics[width=\linewidth]{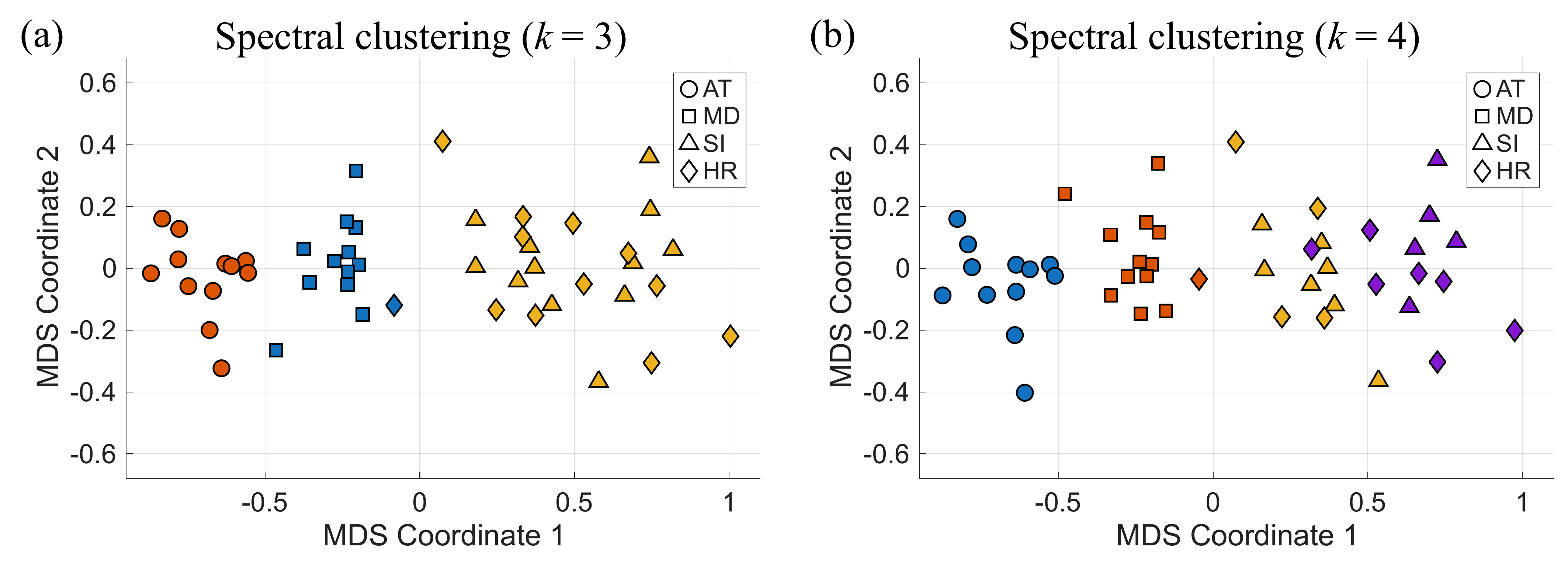}
    \caption{
    \textbf{MDS plots of the honey bee wing clustering results achieved using the spectral clustering method.} (a) The result with $k = 3$ clusters (optimal $\beta^* = 0.81$). (b) The result with $k = 4$ clusters (optimal $\beta^* = 0.92$). Marker shapes indicate geographic labels (AT, MD, SI, HR), and marker colors represent the clustering result obtained by spectral clustering. 
    }
    \label{fig:MDS_comparison}
\end{figure}

Besides, note that in our FDA-QC shape analysis framework, the number of resulting clusters is automatically determined by the community detection method. It is natural to ask whether the outcome will change significantly if other clustering methods or fixed numbers of clusters are used. To examine this, we consider re-running the $\beta$ sweeping procedure using the spectral clustering method with a fixed number of $k = 3$ and $k = 4$ clusters. The resulting optimal parameters for $k = 3$ and $k = 4$ are $\beta^* = 0.81$ and $0.92$ respectively, which are consistent with the optimal $\beta^* = 0.75$ obtained in our original formulation. This again suggests that both the boundary geometry and internal venation features are important for representing the honey bee wing shape variation. Also, as shown in Fig.~\ref{fig:MDS_comparison}, the clustering results achieved by spectral clustering for both $k = 3$ and $k = 4$ are highly consistent with our previous result shown in Fig.~\ref{fig:bee_mds}.

Altogether, the FDA-QC clustering results for European honey bee wing shapes suggest that the proposed fused dissimilarity captures both well-separated regional modes and a more continuous regime of variation, and is highly consistent with the ecology. This experiment demonstrates the applicability of the FDA-QC framework as an interpretable multi-scale descriptor for biological shape analysis.

\section{Conclusion and Discussion} \label{sect:conclusion}

In this work, we have introduced a novel method, FDA-QC, for planar morphometry by combining functional shape data analysis (FDA) with quasi-conformal (QC) mappings. The central idea is to separate two geometric aspects of planar shape variation: boundary deformation, described through the FDA-based elastic registration in the SRVF representation, and interior deformation, described through the QC-based planar mapping. By using the FDA-based boundary correspondence as the boundary condition and solving for a QC map of the entire shape, we obtain a coherent construction that links boundary matching and interior registration. Besides providing a registration between two planar shapes, the FDA-QC method can also be utilized for generating a continuous shape morphing process between two shapes, as well as for quantifying the shape variation between a large number of planar shapes using a shape dissimilarity measure based on both the FDA distance and QC distortions. Altogether, our work provides a unified geometric approach to planar morphometry, connecting boundary registration, interior mapping, shape morphing, and shape comparison within a single framework. As demonstrated by our experiments on plant leaves and insect wings, the method offers a new way for the morphometrics and morphogenesis of planar biological shapes by naturally integrating both the contours and internal structures.

Several limitations remain. The present framework is restricted to simply connected planar domains but not multiply connected planar domains with interior holes. In addition, the combined dissimilarity introduced here is primarily a geometrically motivated quantity, while its finer mathematical properties and statistical behavior remain to be studied more systematically. On the application side, while our method provides geometric evidence of organization within the leaf and wing datasets, the underlying biological mechanisms require further investigation.

There are several natural directions for future work. One is to consider more flexible ways of combining boundary and interior terms, for example through adaptive or spatially varying weights, or by incorporating the tensor field representations~\cite{zhang2021elastic} into our current method. It would also be of interest to integrate the FDA-QC method with statistical inference on shape spaces for a more comprehensive understanding of the developmental variations. Another future direction is to extend the method beyond planar domains to surfaces and volumetric shapes, in which the interactions between the boundary and the internal deformation are expected to be more complex.

\bibliographystyle{ieeetr}
\bibliography{reference}

\end{document}